\def\nk{n_{\rm b}}

\def\Pb{P_{\rm b}}

\def\rfr#1{Equation\,(\ref{#1})}
\def\rfrs#1#2{Equations\,(\ref{#1})-(\ref{#2})}

\def\virg#1{``#1"}

\def\eqi{\begin{equation}}
\def\eqf{\end{equation}}
\def\eqia{\begin{eqnarray}}
\def\eqfa{\end{eqnarray}}

\def\rp#1#2{{#1\over#2}}
\def\lb#1{\label{#1}}

\def\bds#1{\boldsymbol{#1}}

\def\ton#1{\left(#1\right)}
\def\qua#1{\left[#1\right]}
\def\grf#1{\left\{#1\right\}}

\documentclass[onecolumn]{aastex}
\usepackage{morefloats}
\usepackage[title]{appendix}
\usepackage{hyperref}
\usepackage{booktabs}
\usepackage[table,xcdraw]{xcolor}
\usepackage{multirow}
\usepackage{rotating,tabularx}
\usepackage{float}
\usepackage{enumerate}
\usepackage{rotating}
\usepackage[polutonikogreek,english]{babel}
\usepackage{amsmath,starfont,textgreek,w-greek,wasysym}
\usepackage[flushleft]{threeparttable}
\usepackage{amsthm}
\usepackage{amscd,lineno}
\usepackage{amssymb,dsfont}
\usepackage{graphicx,epsfig}
\usepackage{txfonts}
\usepackage{xr-hyper}

\RequirePackage{color}

\newcommand{\emaila}{lorenzo.iorio@libero.it}

\allowdisplaybreaks[1]

\begin{document}

\title{The Short-period S-stars S4711, S62, S4714 and the Lense--Thirring Effect due to the Spin of Sgr A$^\ast$}

\shortauthors{L. Iorio}

\author{Lorenzo Iorio\altaffilmark{1} }
\affil{Ministero dell'Istruzione, dell'Universit\`{a} e della Ricerca
(M.I.U.R.)
\\ Viale Unit\`{a} di Italia 68, I-70125, Bari (BA),
Italy}

\email{\emaila}

\begin{abstract}
Recently, some S-stars (S4711,\,S62,\,S4714) orbiting the supermassive black hole (SMBH) in Sgr A$^\ast$  with short orbital periods ($7.6\,\mathrm{yr}\leq \Pb\leq 12\,\mathrm{yr}$) were discovered. It was suggested that they may be used to measure the general relativistic Lense-Thirring (LT) precessions of their longitudes of ascending node $\Omega$ induced by the SMBH's angular momentum $\bds J_\bullet$. In fact, the proposed numerical estimates hold only in the particular case of a perfect alignment of $\bds J_\bullet$ with the line of sight, which does not seem to be the case. Moreover, also the inclination $I$ and the argument of perinigricon $\omega$ also undergo LT precessions for an arbitrary orientation of $\bds J_\bullet$ in space. We explicitly show the analytical expressions of $\dot I^\mathrm{LT},\,\dot\Omega^\mathrm{LT},\,\omega^\mathrm{LT}$ in terms of the SMBH's spin polar angles $i^\bullet,\,\varepsilon^\bullet$. It turns out that the LT precessions, in arcseconds per year $\ton{^{\prime\prime}\,\mathrm{yr}^{-1}}$, range within
$\left|\dot I^\mathrm{LT}\right| \lesssim 7\,^{\prime\prime}\,\mathrm{yr}^{-1},\,\left|\dot \Omega^\mathrm{LT}\right| \lesssim 9\,^{\prime\prime}\,\mathrm{yr}^{-1},\,\textrm{and}-14\,^{\prime\prime}\,\mathrm{yr}^{-1}\lesssim\dot \omega^\mathrm{LT} \lesssim 13\,^{\prime\prime}\,\mathrm{yr}^{-1}$ for S4714,
$\left|\dot I^\mathrm{LT}\right| \lesssim 5\,^{\prime\prime}\,\mathrm{yr}^{-1},\,\left|\dot \Omega^\mathrm{LT}\right| \lesssim 5\,^{\prime\prime}\,\mathrm{yr}^{-1},\,\textrm{and}\left|\dot \omega^\mathrm{LT}\right| \lesssim 10\,^{\prime\prime}\,\mathrm{yr}^{-1}$ for S62,
and
$\left|\dot I^\mathrm{LT}\right| \lesssim 0.3\,^{\prime\prime}\,\mathrm{yr}^{-1},\,\left|\dot \Omega^\mathrm{LT}\right| \lesssim 0.3\,^{\prime\prime}\,\mathrm{yr}^{-1},\,\textrm{and}\left|\dot\omega^\mathrm{LT}\right| \lesssim 0.7\,^{\prime\prime}\,\mathrm{yr}^{-1}$ for S4711.
For each star, the corresponding values of $i^\bullet_\mathrm{max},\,\varepsilon^\bullet_\mathrm{max}$, and $i^\bullet_\mathrm{min},\,\varepsilon^\bullet_\mathrm{min}$ are determined as well, along with those $i_0^\bullet,\,\varepsilon_0^\bullet$  that cancel the LT precessions.
The LT perinigricon precessions $\dot\omega^\mathrm{LT}$ are overwhelmed by the systematic uncertainties in the Schwarzschild ones due to the current errors in the stars' orbital parameters and the mass of Sgr A$^\ast$ itself.
\end{abstract}

keywords{
Kerr black holes (886); Orbital elements (1177); General relativity (641);
}
\section{Introduction}
Recently, \citet{2020ApJ...899...50P} discovered some main sequence stars revolving around the supermassive black hole (SMBH) allegedly lurking in the Galactic Center (GC) at Sgr A$^\ast$ \citep{2008ApJ...689.1044G,2010RvMP...82.3121G} characterized by very high eccentricities and relatively short orbital periods $\Pb$ with respect to the star S2 \citep{2008ApJ...672L.119M} ($\Pb=16\,\mathrm{yr}$) and other S-stars \citep{2017ApJ...837...30G}; they are listed in Table\,\ref{tavola0}.
\begin{table}[!htb]
\begin{center}
\begin{threeparttable}
\caption{Relevant Orbital Parameters of Some Selected Short-period S-stars Retrieved from Table 2 of \citet{2020ApJ...899...50P}.
}\lb{tavola0}
\begin{tabular*}{\textwidth}{c@{\extracolsep{\fill}}c c c c c c}
\toprule
%
\multirow{2}{*}{}
$\mathrm{Star}$ & $a$                  & $\Pb$               & $e$ & $I$            & $\Omega$ \\
                & $\ton{\mathrm{mpc}}$ & $\ton{\mathrm{yr}}$ &     & $\ton{^\circ}$ & $\ton{^\circ}$ \\
\midrule
%
%
S4714 & $4.079\pm 0.012$ & $12.0\pm 0.3$ & $0.985\pm 0.011$ & $127.70\pm 0.28$ & $129.28\pm 0.63$\\
S62 & $3.588 \pm 0.02$ & $9.9\pm 0.3$ & $0.976\pm 0.01$ & $72.76\pm 5.15$ & $122.61\pm 4.01$ \\
S4711 & $3.002 \pm 0.06$ & $7.6\pm 0.3$ & $0.768\pm 0.030$ & $114.71 \pm 2.92$ & $20.10 \pm 3.72$ \\
\bottomrule
\end{tabular*}
\begin{tablenotes}
\small
\item \textbf{Note.} Here, $a$ is the semimajor axis, in milliparsecs (mpc), $\Pb=2\uppi\sqrt{a^3/\mu_\bullet}$ is the orbital period, in yr  ($\mu_\bullet\doteq GM_\bullet$ is the SMBH's gravitational parameter given by the product of its mass $M_\bullet=\ton{4.1\pm 0.2}\times 10^6\,\mathrm{M}_\odot$ and the Newtonian gravitational constant $G$), $e$ is the eccentricity, $I$ is the inclination of the orbital plane to the plane of the sky, chosen as reference $\grf{x,\,y}$ plane, $\Omega$ is the longitude of the ascending node. Both $I$ and $\Omega$ are measured in degrees $\ton{^\circ}$, while $e$ is dimensionless.
\end{tablenotes}
\end{threeparttable}
\end{center}
\end{table}
\citet{2020A&A...636L...5G} were able to successfully measure the general relativistic Schwarzschild-like precession of the perinigricon $\omega$
\eqi
\dot\omega^\mathrm{Sch} = \rp{3\,\nk\,\mu}{c^2\,a\,\ton{1-e^2}}\lb{GEom}
\eqf
of S2 with a $19\%$ accuracy; in \rfr{GEom}, $c$ is the speed of light in vacuum, $a$ is the semimajor axis, $e$ is the eccentricity, $\nk=2\uppi/\Pb$, and $\mu\doteq GM$ is the primary's gravitational parameter given by the product of its mass $M$ and the Newtonian gravitational constant $G$. Previously, the combined gravitational redshift and relativistic transverse Doppler effect of S2 were detected by \citet{2018A&A...615L..15G} with a $\simeq 15\%$ accuracy. All such gravitational effects,\footnote{The transverse Doppler shift is predicted by the special theory of relativity.} due to the post-Newtonian (pN) static part of the spacetime of the SMBH assumed as nonrotating, are often dubbed in the literature as \virg{gravitoelecric} \citep{Thorne86,2001rfg..conf..121M}, in (formal) analogy with the Coulombian electrostatic field of a point charge.

\citet{2020ApJ...899...50P} explored the possibility of using some of the recently discovered fast-revolving S-stars to measure a peculiar dynamical feature predicted by general relativity as well: the gravitomagnetic Lense--Thirring (LT) orbital precessions induced by the proper angular momentum ${\bds J}$ of the central body \citep{LT18,SoffelHan19} which, in the present case, is the SMBH in Sgr A$^\ast$. They arise from the pN stationary, \virg{gravitomagnetic} \citep{Thorne86,1986hmac.book..103T,1988nznf.conf..573T,2001rfg..conf..121M,2001rsgc.book.....R} component of the gravitational field induced by the mass--energy currents which, for an isolated spinning mass, determine just $\bds J$. For other gravitomagnetic effects, see, e.g., \citet{1986SvPhU..29..215D,2002NCimB.117..743R,2004GReGr..36.2223S,2009SSRv..148...37S}, and references therein. According to the no-hair theorems \citep{1970JMP....11.2580G,1974JMP....15...46H}, the magnitude of the angular momentum of a rotating Kerr BH is \citep{1972ApJ...178..347B,2001ApJ...554L..37M} $J_\bullet = \chi_\mathrm{g}\,M^2_\bullet\,G/c$, where $0\leq\chi_\mathrm{g}\leq 1$.

It is not the first time that the idea of looking at the S-stars to perform such a measurement was put forth in the literature \citep{1998AcA....48..653J,2003ApJ...590L..33L,2007CQGra..24.1775K,2008IAUS..248..466G,2008ApJ...674L..25W,2009ApJ...703.1743P,2010ApJ...720.1303A,2010PhRvD..81f2002M,2011MNRAS.411..453I,2013degn.book.....M,2014RAA....14.1415H,2015ApJ...809..127Z,2016ApJ...818..121P,2016ApJ...827..114Y,2017ApJ...834..198Z,2018MNRAS.476.3600W,2019GReGr..51..137P}.
Quite recently, \citet{2020arXiv200811734F} used the newly discovered S-stars which are the subject of this study to infer upper bounds on $\chi_\mathrm{g}$.
The main issue with the analysis by \citet{2020ApJ...899...50P} is that they looked solely at the node precession $\dot\Omega^\mathrm{LT}$ by means of a formula which holds only in a particular coordinate system whose reference $z$ axis is aligned with the SMBH's spin axis. Actually, the orientation in the space of the unit vector ${\bds{\hat{J}}}_\bullet$ of Sgr A$^\ast$ is not known a priori, being, currently,  a lingering uncertainty on it. Moreover, for an arbitrary orientation of the primary's spin axis in space, it turns out that also the orbital inclination $I$ and the argument of pericenter $\omega$ undergo secular LT precessions. Finally, it is important to remark that $\dot I^\mathrm{LT}$, $\dot\Omega^\mathrm{LT}$, and $\dot\omega^\mathrm{LT}$ depend explicitly on the direction of ${\bds{\hat{J}}}$ in space. Thus, the validity of the figures quoted in Table 3 of \citet{2020ApJ...899...50P} for $\dot\Omega^\mathrm{LT}$ is limited since they hold only for the particular case in which the SMBH's spin axis is aligned with the line of sight, assumed as a reference $z$ axis, which, in general, may not be the case. The same drawbacks are common also to \citet{2020arXiv200811734F}.
For performed attempts to somehow constrain ${\bds{\hat{J}}}_\bullet$ of Sgr A$^\ast$ with different nondynamical approaches, see \citet{2007ApJ...662L..15F,2007A&A...473..707M,2009ApJ...697...45B,2011ApJ...735..110B,2012ApJ...755..133S,2016ApJ...827..114Y}, and references therein; it can be seen that the polar angles $i^\bullet,\,\varepsilon^\bullet$ determining the orientation of ${\hat{\bds{J}}}_\bullet$ in space are, in fact, still poorly constrained. In any case, it seems that it is far from being aligned with the line of sight. Indeed, according to, e.g., \citet{2007A&A...473..707M}, who used polarimetric observations of the near-infrared emission of Sgr A$^\ast$, it is $i^\bullet\simeq 55^\circ$. \citet{2007ApJ...662L..15F} obtained $i^\bullet\simeq 77^\circ$ on the basis of their fit of a simulated Rossby wave-induced spiral pattern in the SMBH's accretion disk to the X-ray lightcurve detected with XMM-Newton. \citet{2012ApJ...755..133S} yielded the range $42^\circ\lesssim i^\bullet \lesssim 75^\circ$ by comparing polarized submillimeter infrared observations with spectra computed using three-dimensional general relativistic  magnetohydrodynamical simulations.  Methods based on gravitational lensing for determining the SMBH's spin direction independently of orbital dynamics were outlined, e.g., in \citet{2017PTEP.2017e3E02S}. \citet{2004ApJ...611..996T} investigated the possibility of measuring, among other things, $i^\bullet$ from the shape and position of the BH's shadow under certain assumptions.

Here, we make a step forward by providing explicit analytical expressions for the LT precessions of $I,\,\Omega,\,\textrm{and}\omega$ in terms of  $i,\,\varepsilon$. Furthermore, moving to the Sgr A$^\ast$ scenario, we plot them as functions of $i^\bullet,\,\varepsilon^\bullet$ for some selected S-stars among those taken into account by \citet{2020ApJ...899...50P} by calculating also the values $i^\bullet_0,\,\varepsilon^\bullet_0$ for which the LT precessions vanish along with those, $i^\bullet_\mathrm{max},\,\varepsilon^\bullet_\mathrm{max}$ and $i^\bullet_\mathrm{min},\,\varepsilon^\bullet_\mathrm{min}$, yielding their maxima and the minima. We look also at the systematic uncertainty affecting the Schwarzschild perinigricon precessions of the considered S-stars  due to the errors in the key physical and orbital parameters entering \rfr{GEom}. Indeed, this represents a major source of systematic uncertainty in the much smaller LT perinigricon precessions.
\section{The Lense--Thirring Orbital Precessions for an Arbitrary Orientation of the SMBH's Spin in Space}
The LT orbital precessions for an arbitrary orientation of the angular momentum of the primary in space were worked out in the literature with a variety of analytical techniques encompassing different parameterizations for both the test particle's orbit and the spin itself
\citep[see,\,e.g.,][]{1975PhRvD..12..329B,1988NCimB.101..127D,1992PhRvD..45.1840D,1999ApJ...514..388W,2007CQGra..24.1775K,2008ApJ...674L..25W,2017EPJC...77..439I}.
According to \citet{2017EPJC...77..439I}, the gravitomagnetic precessions of the inclination, the node and the periastron are
\begin{align}
\dot I \lb{Irate}& = \rp{2\,G\,J\,\bds{\hat{J}}\bds\cdot\bds{\hat{l}}}{c^2\,a^3\,\ton{1-e^2}^{3/2}}, \\ \nonumber\\
\dot \Omega \lb{Orate}& = \rp{2\,G\,J\,\csc I\,\bds{\hat{J}}\bds\cdot\bds{\hat{m}}}{c^2\,a^3\,\ton{1-e^2}^{3/2}}, \\ \nonumber\\
\dot \omega \lb{orate}& = -\rp{2\,G\,J\,\bds{\hat{J}}\bds\cdot\ton{2\,\bds{\hat{h}} + \cot I\bds{\hat{m}}}}{c^2\,a^3\,\ton{1-e^2}^{3/2}},
\end{align}
where  $\textcolor{black}{\bds{\hat{l}}}=\grf{\cos\Omega,\,\sin\Omega,\,0}$ is the unit vector directed in the reference $\grf{x,\,y}$ plane along the line of the nodes toward the ascending node, $\bds{\hat{m}}=\grf{-\cos I\,\sin\Omega,\,\cos I\,\cos\Omega,\,\sin I}$ is the unit vector directed in the orbital plane transversely to $\bds{\hat{l}}$, and $\bds{\hat{h}}=\grf{\sin I\,\sin\Omega,\,-\sin I\,\cos\Omega,\,\cos I}$ is the unit vector directed along the orbital angular momentum perpendicularly to the orbital plane in such a way that $\bds{\hat{l}},\,\bds{\hat{m}},\bds{\hat{h}}$ are a right-handed triad of unit vectors.
By adopting for $\bds{\hat{J}}$ the following parameterization
\begin{align}
{\hat{J}}_x \lb{Jx}& = \sin i\,\cos\varepsilon, \\ \nonumber \\
{\hat{J}}_y \lb{Jy}& = \sin i\,\sin\varepsilon, \\ \nonumber \\
{\hat{J}}_z \lb{Jz}& = \cos i,
\end{align}
\rfrs{Irate}{orate} can be cast into the form
\begin{align}
\dot I^\mathrm{LT} \lb{dIdt}& =\rp{2\,G\,J\,\sin i\,\cos\zeta}{c^2\,a^3\,\ton{1-e^2}^{3/2}}, \\ \nonumber \\
\dot \Omega^\mathrm{LT} \lb{dOdt}& =\rp{2\,G\,J\,\ton{\cos i +\cot I\,\sin i\,\sin\zeta}}{c^2\,a^3\,\ton{1-e^2}^{3/2}}, \\ \nonumber \\
\dot\omega^\mathrm{LT} \lb{dodt}& = \rp{G\,J\,\qua{-6\,\cos I\,\cos i +\ton{1 -3 \cos 2 I}\,\csc I\,\sin i\,\sin\zeta}}{c^2\,a^3\,\ton{1-e^2}^{3/2}}.
\end{align}
They show that the LT precessions depend explicitly on the absolute orientation of the primary's spin axis and of the test particle's orbital plane in space through $i$ and $I$, respectively, and on the relative orientation of the primary's spin axis and the test particle's orbit through the angle $\zeta\doteq\varepsilon-\Omega$. As such, they can even vanish for given combinations of $i,\,I,$ and $\zeta$.

\begin{figure}[!htb]
\begin{center}
\centerline{
\vbox{
\begin{tabular*}{\textwidth}{c@{\extracolsep{\fill}}c}
\epsfysize= 6.0 cm\epsfbox{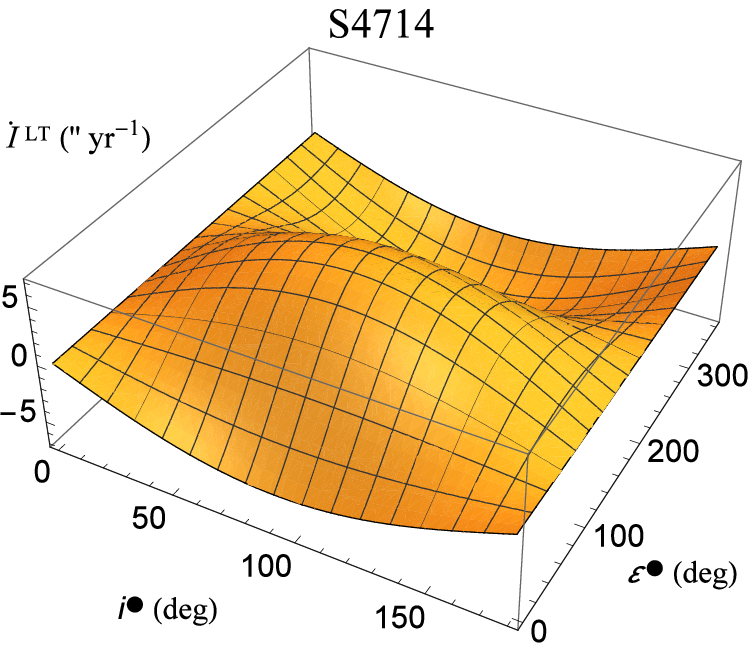}\\
\epsfysize= 6.0 cm\epsfbox{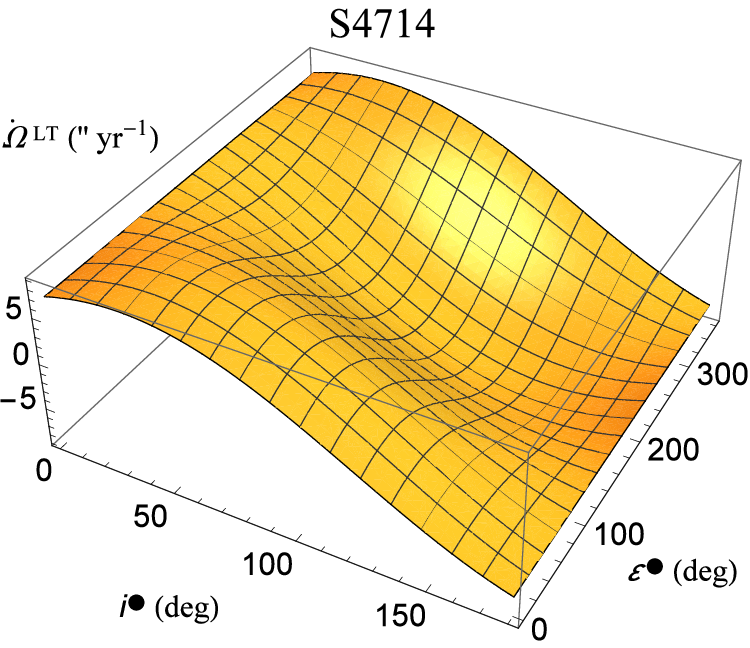}\\
\epsfysize= 6.0 cm\epsfbox{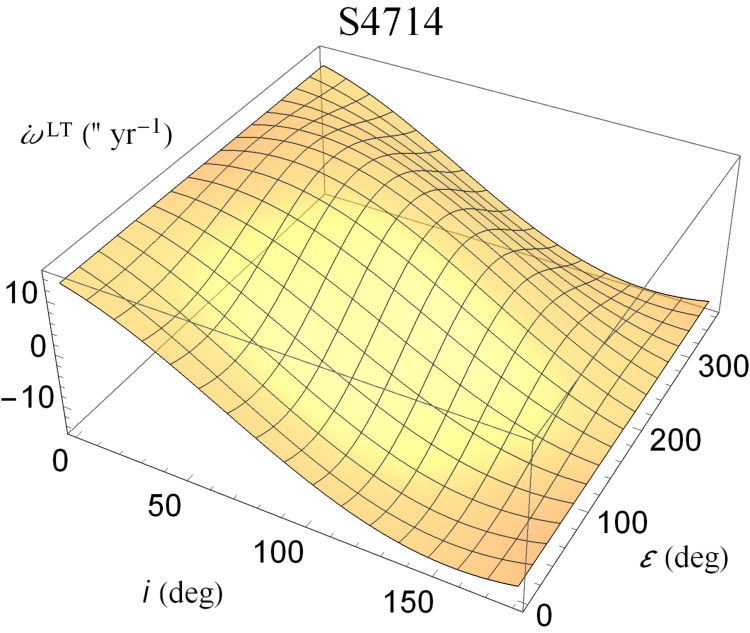}\\
\end{tabular*}
}
}
\caption{LT precessions, in arcseconds per year, of the S-star S4714 according to \rfrs{dIdt}{dOdt} plotted as functions of $i^\bullet,\,\varepsilon^\bullet$. For  $J_\bullet=\chi_\mathrm{g}\,M_\bullet^2\,G/c$, the values  $\chi_\mathrm{g}=0.5,\,M_\bullet=4.1\times 10^6\,\mathrm{M}_\odot$ \citep{2020ApJ...899...50P} were adopted.}\label{figura1}
\end{center}
\end{figure}

\begin{figure}[!htb]
\begin{center}
\centerline{
\vbox{
\begin{tabular*}{\textwidth}{c@{\extracolsep{\fill}}c}
\epsfysize= 6.0 cm\epsfbox{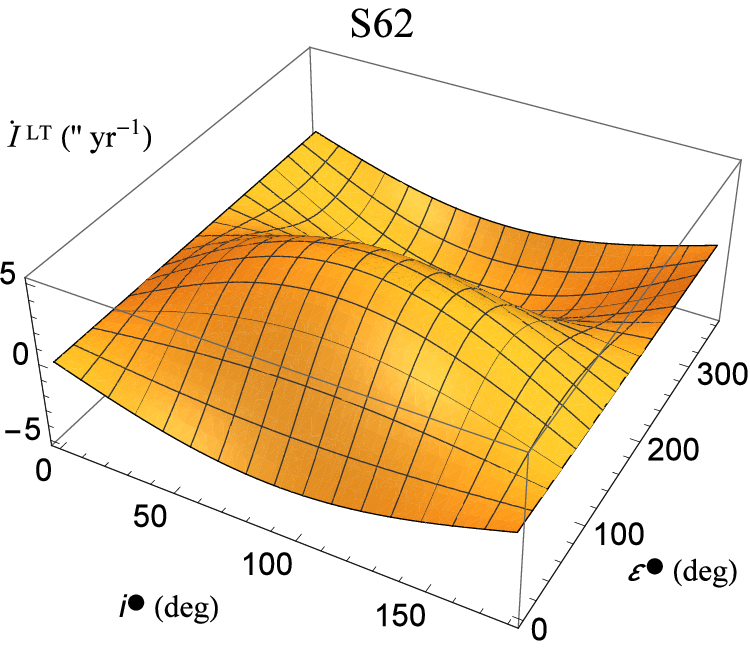}\\
\epsfysize= 6.0 cm\epsfbox{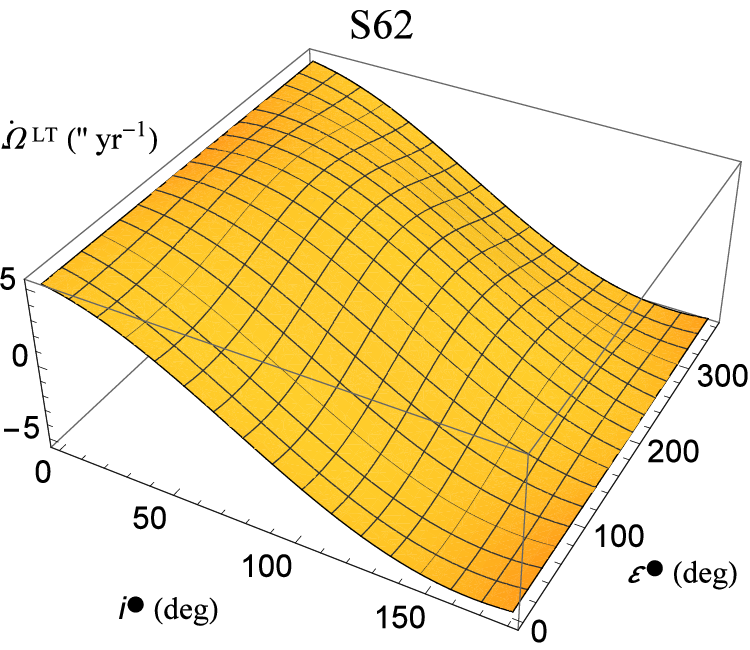}\\
\epsfysize= 6.0 cm\epsfbox{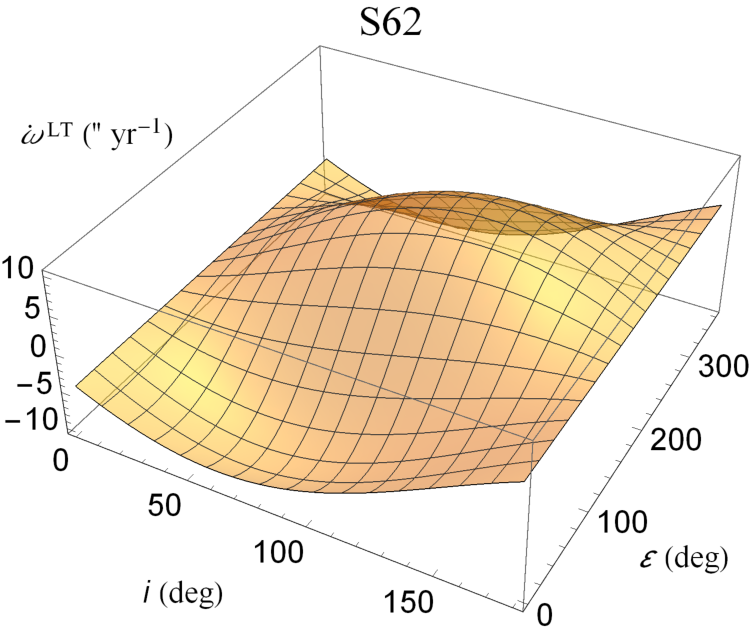}\\
\end{tabular*}
}
}
\caption{LT precessions, in arcseconds per year, of the S-star S62 according to \rfrs{dIdt}{dOdt} plotted as functions of $i^\bullet,\,\varepsilon^\bullet$. For  $J_\bullet=\chi_\mathrm{g}\,M_\bullet^2\,G/c$, the values  $\chi_\mathrm{g}=0.5,\,M_\bullet=4.1\times 10^6\,\mathrm{M}_\odot$ \citep{2020ApJ...899...50P} were adopted.}\label{figura2}
\end{center}
\end{figure}

\begin{figure}[!htb]
\begin{center}
\centerline{
\vbox{
\begin{tabular*}{\textwidth}{c@{\extracolsep{\fill}}c}
\epsfysize= 6.0 cm\epsfbox{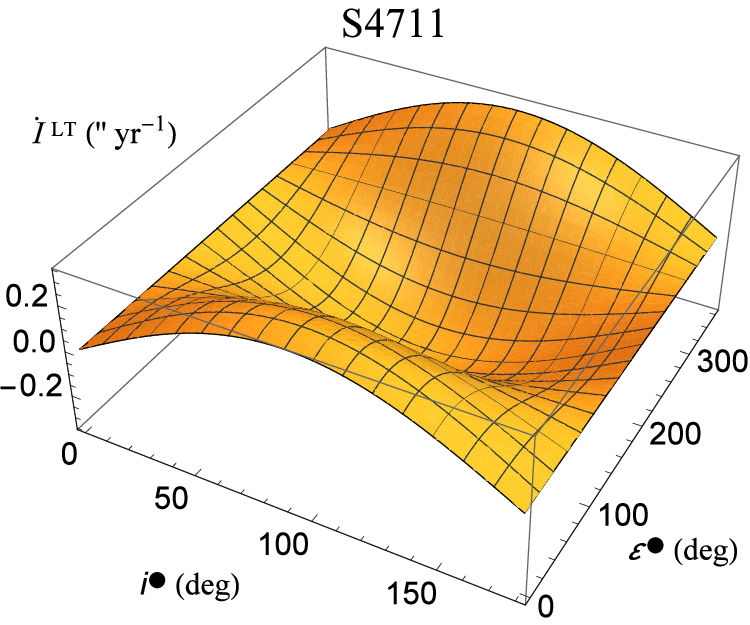}\\
\epsfysize= 6.0 cm\epsfbox{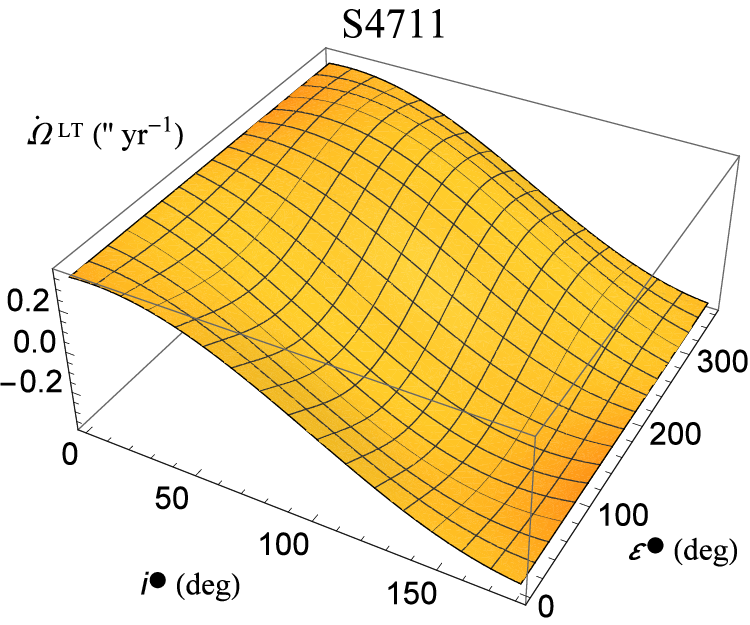}\\
\epsfysize= 6.0 cm\epsfbox{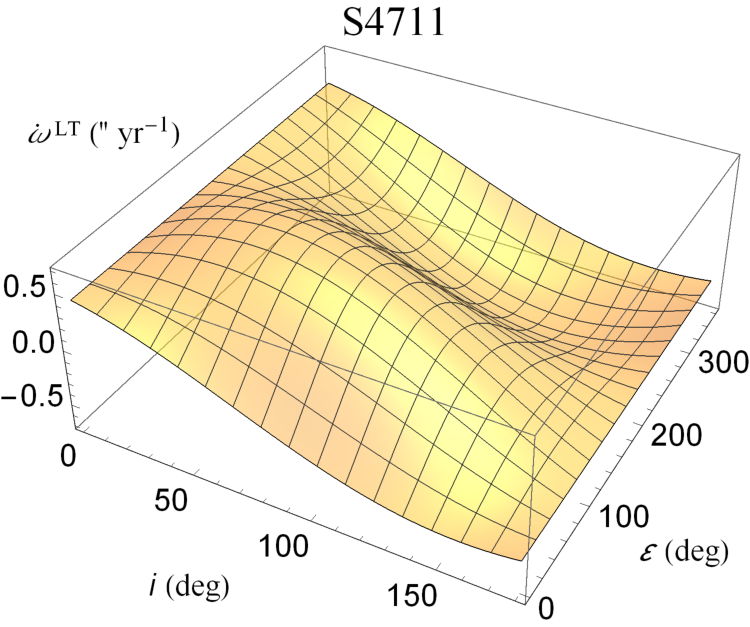}\\
\end{tabular*}
}
}
\caption{LT precessions, in arcseconds per year, of the S-star S4711 according to \rfrs{dIdt}{dOdt} plotted as functions of $i^\bullet,\,\varepsilon^\bullet$. For  $J_\bullet=\chi_\mathrm{g}\,M_\bullet^2\,G/c$, the values  $\chi_\mathrm{g}=0.5,\,M_\bullet=4.1\times 10^6\,\mathrm{M}_\odot$ \citep{2020ApJ...899...50P} were adopted.}\label{figura3}
\end{center}
\end{figure}

In the case of the S-stars orbiting Sgr A$^\ast$, Figures\,\ref{figura1}-\ref{figura3}, produced by assuming $\chi_\mathrm{g}=0.5,\,M_\bullet=4.1\times 10^6\,\mathrm{M}_\odot$ \citep{2020ApJ...899...50P}, depict the plots of
$\dot I^\mathrm{LT}\ton{i^\bullet,\,\varepsilon^\bullet}$,
$\dot\Omega^\mathrm{LT}\ton{i^\bullet,\,\varepsilon^\bullet}$, and
$\dot\omega^\mathrm{LT}\ton{i^\bullet,\,\varepsilon^\bullet}$,
in arcseconds per year ($^{\prime\prime}\,\mathrm{yr}^{-1}$),  for S4714, S62, and S4711 whose relevant orbital parameters are listed in Table\,\ref{tavola0}.
It can be noted that, for certain values of $i^\bullet,\,\varepsilon^\bullet$, they vanish, possibly being  both positive and negative values.

Tables\,\ref{tavola1}-\ref{tavola3} display the values $i^\bullet_0,\,\varepsilon^\bullet_0$  for which the LT precessions vanish, and those corresponding to their minima and maxima.
\begin{table}[!htb]
\begin{center}
\begin{threeparttable}
\caption{Maximum and Minimum Nominal Values, in Arcseconds per Year, of the LT Rate of Change of the Inclination $I$ of Some Selected  Short-period S-stars \citep{2020ApJ...899...50P} Along with the Corresponding Values, in Degrees, of the SMBH's Spin Axis Polar Angles $0^\circ\leq i^\bullet\leq 180^\circ,\,0^\circ\leq\varepsilon^\bullet\leq 360^\circ$.
}\lb{tavola1}
\begin{tabular*}{\textwidth}{c@{\extracolsep{\fill}}c c c c c c c}
\toprule
%
\multirow{2}{*}{}
 &
 $\dot I_\mathrm{max}^\mathrm{LT}$
 &
 $i^\bullet_\mathrm{max}$
 &
 $\varepsilon^\bullet_\mathrm{max}$
 &
 $\dot I_\mathrm{min}^\mathrm{LT}$
 &
 $i^\bullet_\mathrm{min}$
 &
 $\varepsilon^\bullet_\mathrm{min}$ \\

 &
 $\ton{^{\prime\prime}\,\mathrm{yr}^{-1}}$
 &
 $\ton{^\circ}$ & $\ton{^\circ}$
 &
 $\ton{^{\prime\prime}\,\mathrm{yr}^{-1}}$
 &
 $\ton{^\circ}$
 &
 $\ton{^\circ}$\\
\midrule
%
%
S4714 & $7.04$ & $90$ & $129.3$ & $-7.04$ & $90$ & $309.3$ \\
S62 & $5.15$ & $90$ & $122.6$ & $-5.15$ & $90$ & $302.6$ \\
S4711 & $0.34$ & $90$ & $20.1$ & $-0.34$ & $90$ & $200.1$ \\
\bottomrule
\end{tabular*}
\begin{tablenotes}
\small
\item \textbf{Note.} For  $J_\bullet=\chi_\mathrm{g}\,M_\bullet^2\,G/c$, the values  $\chi_\mathrm{g}=0.5,\,M_\bullet=4.1\times 10^6\,\mathrm{M}_\odot$ \citep{2020ApJ...899...50P} were adopted. According to \rfr{dIdt}, $\dot I^\mathrm{LT}=0^\circ$ for the spin axis aligned with the direction of the line of sight, i.e. for $i^\bullet_0=0^\circ$ or $i^\bullet_0=180^\circ$, or for $\varepsilon^\bullet_0=\Omega\pm 90^\circ$.
\end{tablenotes}
\end{threeparttable}
\end{center}
\end{table}
\begin{table}[!htb]
\begin{center}
\begin{threeparttable}
\caption{Maximum and Minimum Values, in Arcseconds per Year, of the LT Rate of Change of the Node $\Omega$ of Some Selected  Short-period S-stars \citep{2020ApJ...899...50P} Along with the Corresponding Values, in Degrees, of the SMBH's Spin Axis Polar Angles $0^\circ\leq i^\bullet\leq 180^\circ,\,0^\circ\leq\varepsilon^\bullet\leq 360^\circ$.
}\lb{tavola2}
\begin{tabular*}{\textwidth}{c@{\extracolsep{\fill}}c c c c c c c c c}
\toprule
%
\multirow{2}{*}{}
&
$\dot \Omega_\mathrm{max}^\mathrm{LT}$
&
$i^\bullet_\mathrm{max}$
&
$\varepsilon^\bullet_\mathrm{max}$
&
$\dot \Omega_\mathrm{min}^\mathrm{LT}$
&
$i^\bullet_\mathrm{min}$
&
$\varepsilon^\bullet_\mathrm{min}$
&
$i^\bullet_0$
&
$\varepsilon^\bullet_0$ \\
&
$\ton{^{\prime\prime}\,\mathrm{yr}^{-1}}$
&
$\ton{^\circ}$
&
$\ton{^\circ}$
&
$\ton{^{\prime\prime}\,\mathrm{yr}^{-1}}$
&
$\ton{^\circ}$
&
$\ton{^\circ}$
&
$\ton{^\circ}$
&
$\ton{^\circ}$
\\
\midrule
%
%
S4714 & $8.9$ & $37.7$ & $39.3$ & $-8.9$ & $142.3$ & $219.3$ & $77.9$ & $293.3$\\
S62 & $5.15$ & $0$ & $32.7$ & $-5.32$ & $165.3$ & $360$ & $96.0$ & $142.4$\\
S4711 & $0.35$ & $8.9$ & $0$ & $-0.34$ & $180$ & $359.9$ & $105.5$ & $237.3$\\
\bottomrule
\end{tabular*}
\begin{tablenotes}
\small
\item \textbf{Note.}  For  $J_\bullet=\chi_\mathrm{g}\,M_\bullet^2\,G/c$, the values $\chi_\mathrm{g}=0.5,\,M_\bullet=4.1\times 10^6\,\mathrm{M}_\odot$ \citep{2020ApJ...899...50P} were adopted. The values $i^\bullet_0,\,\varepsilon^\bullet_0$ yield $\dot \Omega^\mathrm{LT}=0^\circ$.
\end{tablenotes}
\end{threeparttable}
\end{center}
\end{table}
\begin{table}[!htb]
\begin{center}
\begin{threeparttable}
\caption{Maximum and Minimum Values, in Arcseconds per Year, of the LT Rate of Change of the Perinigricon $\omega$ of Some Selected  Short-period S-stars \citep{2020ApJ...899...50P} Along with the Corresponding Values, in Degrees, of the SMBH's Spin Axis Polar Angles $0^\circ\leq i^\bullet\leq 180^\circ,\,0^\circ\leq\varepsilon^\bullet\leq 360^\circ$.
}\lb{tavola3}
\begin{tabular*}{\textwidth}{c@{\extracolsep{\fill}}c c c c c c c c c}
\toprule
%
\multirow{2}{*}{}
&
$\dot\omega_\mathrm{max}^\mathrm{LT}$
&
$i^\bullet_\mathrm{max}$
&
$\varepsilon^\bullet_\mathrm{max}$
&
$\dot\omega_\mathrm{min}^\mathrm{LT}$
&
$i^\bullet_\mathrm{min}$
&
$\varepsilon^\bullet_\mathrm{min}$
&
$i^\bullet_0$
&
$\varepsilon^\bullet_0$ \\
&
$\ton{^{\prime\prime}\,\mathrm{yr}^{-1}}$
&
$\ton{^\circ}$
&
$\ton{^\circ}$
&
$\ton{^{\prime\prime}\,\mathrm{yr}^{-1}}$
&
$\ton{^\circ}$
&
$\ton{^\circ}$
&
$\ton{^\circ}$
&
$\ton{^\circ}$
\\
\midrule
%
%
S4714 & $12.9$ & $0$ & $39.3$ & $-14.1$ & $154.9$ & $360$ & $102.5$ & $150.7$ \\
S62 & $10.4$ & $116.0$ & $212.6$ & $-10.4$ & $63.9$ & $32.6$ & $55.7$ & $283.1$ \\
S4711 & $0.70$ & $52.3$ & $110.1$ & $-0.70$ & $127.6$ & $290.1$ & $134.2$ & $68.8$ \\
\bottomrule
\end{tabular*}
\begin{tablenotes}
\small
\item \textbf{Note.} For  $J_\bullet=\chi_\mathrm{g}\,M_\bullet^2\,G/c$, the values $\chi_\mathrm{g}=0.5,\,M_\bullet=4.1\times 10^6\,\mathrm{M}_\odot$ \citep{2020ApJ...899...50P} were adopted. The values $i^\bullet_0,\,\varepsilon^\bullet_0$ yield $\dot\omega^\mathrm{LT}=0^\circ$.
\end{tablenotes}
\end{threeparttable}
\end{center}
\end{table}
The largest effects occur for the perinigricon $\omega$, while the smallest ones are those of the inclination $I$. The star S4714 exhibits the most relevant gravitomagnetic precessions ranging from $14\,^{\prime\prime}\,\mathrm{yr}^{-1}$ for $\omega$ to $7\,^{\prime\prime}\,\mathrm{yr}^{-1}$ for $I$, while those for S4711 are at the subarcsecond per year level. It should be noted that, in general, $i^\bullet_0,\,\varepsilon^\bullet_0,\,i^\bullet_\mathrm{max},\,\varepsilon^\bullet_\mathrm{max},\,i^\bullet_\mathrm{min},\,\varepsilon^\bullet_\mathrm{min}$ are different from one precession to another for each star.

The fact that the inclination $I$ and the node $\Omega$ also exhibit LT precession is, in principle, quite important since using only the perinigricon $\omega$ may be an issue because of the uncertainties plaguing its larger Schwarzschild-like rate of change of \rfr{GEom}. The propagation of the errors in $a,\,e$, and $\mu_\bullet$, listed in Table\,\ref{tavola0}, yield the mismodeling quoted in Table\,\ref{tavola4}.
\begin{table}[!htb]
\begin{center}
\begin{threeparttable}
\caption{Nominal Schwarzschild Perinigricon Precessions $\dot\omega^\mathrm{Sch}$, in Arcseconds per Year, of the S-stars S4714, S62, S4711 and Related Uncertainties $\delta\dot\omega^\mathrm{Sch}_{\upsigma_x}$, in Arcseconds per Year, Due to the Errors in $x=a,\,e,\,\mu_\bullet$ According to Table\,\ref{tavola0}.
}\lb{tavola4}
\begin{tabular*}{\textwidth}{c@{\extracolsep{\fill}}ccccc}
\toprule
%
\multirow{2}{*}{}
&
$\dot\omega^\mathrm{Sch}$
&
$\delta\dot\omega_{\upsigma_a}^\mathrm{Sch}$
&
$\delta\dot\omega_{\upsigma_e}^\mathrm{Sch}$
&
$\delta\dot\omega_{\upsigma_{\mu_\bullet}}^\mathrm{Sch}$\\
&
$\ton{^{\prime\prime}\,\mathrm{yr}^{-1}}$
&
$\ton{^{\prime\prime}\,\mathrm{yr}^{-1}}$
&
$\ton{^{\prime\prime}\,\mathrm{yr}^{-1}}$
&
$\ton{^{\prime\prime}\,\mathrm{yr}^{-1}}$ \\
\midrule
%
%
S4714 & $521.1$ & $3.8$ & $379.2$ & $38.1$\\
S62 &  $450.8$ & $6.3$ & $185.6$ & $33.0$ \\
S4711 & $81.4$ & $4.0$ & $9.1$ & $5.9$ \\
\bottomrule
\end{tabular*}
\end{threeparttable}
\end{center}
\end{table}
It turns out that the main source of systematic bias is the eccentricity $e$; the corresponding mismodeled parts of $\dot\omega^\mathrm{Sch}$ largely overwhelm the LT precessions for all the S-stars considered. Also, the mass of the SMBH should be known with a much better accuracy since its uncertainty yields systematic errors in $\dot\omega^\mathrm{Sch}$ larger than $\dot\omega^\mathrm{LT}$. Instead, $I$ and $\Omega$ do not undergo huge competing Schwarzschild precessions.

In principle, also other competing dynamical effects may bias a potential detection of the LT orbital precessions because of their mismodeling or unmodeling. One of them, generally impacting $I,\,\Omega,\,\omega$, is the Newtonian 3-body perturbation on the target S-star induced by other distant stars for which analytical expressions exist in the literature either for their double average over both $\Pb$ and the longer orbital period of the perturber as well; see, e.g., \citet{2020ApJ...889..152I}. The assessment of the systematic uncertainties in the LT measurement depends on the mismodeling in the mass $m$ of the third body and in its orbital parameters. Another orbital perturbation to be taken into account, in principle, is that caused by a possible continuous mass distribution, baryonic or not, inside the orbit of the S-star of interest. If its density $\varrho$ is spherically symmetric, it affects only the perinigricon $\omega$ \citep{2013Galax...1....6I}. Dedicated analyses, including, e.g., simultaneous numerical integrations of the equations of motion of given target stars and other disturbing ones for different values of the masses of the latter are required; they are beyond the scope of the present paper.

\section{Summary and conclusions}
In principle, the inclination $I$, the longitude of the ascending node $\Omega$ and the argument of perinigricon $\omega$ of some of the recently discovered
fast-revolving S-stars can be used to measure their  gravitomagnetic LT effect induced by the angular momentum $J_\bullet$ of the SMBH in the GC at Sgr A$^\ast$. Such general relativistic orbital precessions
depend explicitly on the orientation of ${\bds{\hat{J}}}_\bullet$ in space, and in general, are all nonzero. Indeed, present-day constraints on its polar angles $i^\bullet,\,\varepsilon^\bullet$, despite being still rather weak, are accurate enough to exclude that the SMBH's spin is aligned with the line of sight; thus, also $\dot I^\mathrm{LT}$ which is proportional to $\sin i^\bullet$, is likely nonzero. We calculated the maxima and minima for $\dot I^\mathrm{LT},\,\dot\Omega^\mathrm{LT},\,\dot\omega^\mathrm{LT}$ along with the corresponding values $i^\bullet_\mathrm{max},\,\varepsilon^\bullet_\mathrm{max}$ and $i^\bullet_\mathrm{min},\,\varepsilon^\bullet_\mathrm{min}$ for each star.
We obtained
$\left|\dot I^\mathrm{LT}\right| \lesssim 7\,^{\prime\prime}\,\mathrm{yr}^{-1},\,\left|\dot \Omega^\mathrm{LT}\right| \lesssim 9\,^{\prime\prime}\,\mathrm{yr}^{-1},\,\mathrm{and}-14\,^{\prime\prime}\,\mathrm{yr}^{-1}\lesssim\dot \omega^\mathrm{LT} \lesssim 13\,^{\prime\prime}\,\mathrm{yr}^{-1}$ for S4714,
$\left|\dot I^\mathrm{LT}\right| \lesssim 5\,^{\prime\prime}\,\mathrm{yr}^{-1},\,\left|\dot \Omega^\mathrm{LT}\right| \lesssim 5\,^{\prime\prime}\,\mathrm{yr}^{-1},\,\mathrm{and}\left|\dot \omega^\mathrm{LT}\right| \lesssim 10\,^{\prime\prime}\,\mathrm{yr}^{-1}$ for S62,
and
$\left|\dot I^\mathrm{LT}\right| \lesssim 0.3\,^{\prime\prime}\,\mathrm{yr}^{-1},\,\left|\dot \Omega^\mathrm{LT}\right| \lesssim 0.3\,^{\prime\prime}\,\mathrm{yr}^{-1},\,\mathrm{and}\left|\dot\omega^\mathrm{LT}\right| \lesssim 0.7\,^{\prime\prime}\,\mathrm{yr}^{-1}$ for S4711.
We also determined the values $i^\bullet_0,\,\varepsilon_0^\bullet$ for which the LT precessions vanish.

Despite the largest values occurring for the arguments of perinigricon $\omega$ of S4714 and S62, being at the level of $\simeq 10\,^{\prime\prime}\,\mathrm{yr}^{-1}$, the use of such orbital elements is made problematic by the systematic bias arising from the mismodeling in their much larger Schwarzschild precessions due to the current errors in the stellar orbital elements, especially in the eccentricity $e$, and the mass of Sgr A$^\ast$ itself. Suffice to say that while the nominal gravitoelectric 1pN rates  are $\dot\omega^\mathrm{Sch}=521\,^{\prime\prime}\,\mathrm{yr}^{-1}$ (S4714) and $\dot\omega^\mathrm{Sch}=451\,^{\prime\prime}\,\mathrm{yr}^{-1}$ (S62), the mismodeling due to the current errors in $e$ are as large as $\delta\dot\omega_{\upsigma_e}^\mathrm{Sch}=379\,^{\prime\prime}\,\mathrm{yr}^{-1}$  and $\delta\dot\omega_{\upsigma_e}^\mathrm{Sch}=185\,^{\prime\prime}\,\mathrm{yr}^{-1}$, respectively, while those due to the SMBH's mass uncertainty amount to $\delta\dot\omega^\mathrm{Sch}_{\upsigma_{\mu_\bullet}}=38\,^{\prime\prime}\,\mathrm{yr}^{-1}$  and $\delta\dot\omega^\mathrm{Sch}_{\upsigma_{\mu_\bullet}}=33\,^{\prime\prime}\,\mathrm{yr}^{-1}$, respectively.

Other potentially relevant sources of systematic errors, to be analyzed in further dedicated studies, are the classical competing orbital precessions due to other distant stars acting as pointlike disturbing bodies affecting, in general, $I,\,\Omega$, and $\omega$, and to possible continuous mass distribution, baryonic or not, around the SMBH which, if spherically symmetric, perturbs only the perinigricon.

%
%
%
\bibliography{Sstarsbib}{}

\end{document}